\documentclass[twocolumn,aps]{revtex4-2}
\usepackage{amsmath}
\usepackage{amssymb}
\usepackage{lipsum}
\usepackage{graphicx}
\usepackage{dcolumn}
\usepackage{bm}
\usepackage{xcolor}
\usepackage{epstopdf}

\begin{document}

\preprint{Phys.Rev.B}

\title{ Interaction-Induced Magnetotransport in a 2D Dirac–Heavy Hole Hybrid Band System}

\author{G. M. Gusev,$^1$  A. D. Levin,$^1$ , V. A. Chitta,$^1$
 Z. D. Kvon,$^{2,3}$ and  N. N. Mikhailov$^{2,3}$}

\affiliation{$^1$Instituto de F\'{\i}sica da Universidade de S\~ao
Paulo, 135960-170, S\~ao Paulo, SP, Brazil}
\affiliation{$^2$Institute of Semiconductor Physics, Novosibirsk
630090, Russia}
\affiliation{$^3$Novosibirsk State University, Novosibirsk 630090,
Russia}

\date{\today}
\begin{abstract}
 While electron-electron (e-e) interactions are known to influence resistivity in non-Galilean invariant two-dimensional (2D) systems, their effect on magnetotransport is not fully understood. Conventional models for simple bands often predict a vanishing magnetoresistivity from e-e interactions alone. In this work, we investigate magnetotransport in a gapless 6.3 nm HgTe quantum well, a hybrid 2D band system that hosts coexisting holes with both linear (Dirac-like) and parabolic energy bands. Focusing on the high-temperature regime where particle-particle collisions dominate scattering, we observe significant corrections to both the magnetoresistivity and the Hall effect. The high-temperature transport coefficients are in good agreement with the theoretical model describing transport in massive–massless fermion mixtures governed by a frictional mechanism and intervalley scattering. Our findings provide strong experimental validation for this theoretical framework, demonstrating that collisions between particles with different dispersions are a key mechanism governing magnetotransport in hybrid-band semimetals.
\end{abstract}
\maketitle
\section{Introduction}
The question of how electron-electron (e-e) interactions affect resistivity in non-Galilean-invariant systems has long been debated \cite{pal} and has recently gained significant attention due to the discovery and study of unconventional Dirac materials. The intricate electronic structure of these systems - featuring multiple bands with distinct dispersions and topologies - requires moving beyond conventional transport theories. Standard models predict that e-e scattering should not contribute to resistivity in systems with parabolic bands lacking Umklapp processes, whereas non-Galilean-invariant systems are expected to exhibit the characteristic $T^2$ dependence due to e-e collisions \cite{pal, kovalev}.

While the temperature dependence of resistivity has been thoroughly investigated, the role of e-e interactions in magnetoresistance remains largely unexplored. Notably, any scattering mechanism, including e-e collisions, must produce vanishing magnetoresistance \cite{ziman}. This occurs because the Lorentz force is compensated by the electric field generated by the Hall effect. In the simplest scenario, magnetoresistance emerges in two-component systems with distinct transport properties \cite{ziman, zaremba, fletcher, mamani} or in compensated semimetals where electrons and holes coexist \cite{kvon}. However, the influence of e-e scattering has either been overlooked or assumed to have no effect on magnetotransport properties.

Several attempts have been made within the framework of hydrodynamic theory, where electron-electron collisions play a central role. Hydrodynamic electron flow is expected to dominate transport phenomena when the mean free path for e-e collisions ($l_{ee}$) is much shorter than the mean free path associated with impurity and phonon scattering ($l$). Most predictions have focused on narrow channels, where the electron flow exhibits a Poiseuille-like velocity profile \cite{narozhny,fritz}. Consequently, transport properties in such systems are highly sensitive to geometric constraints and intrinsic Fermi liquid characteristics, such as electron viscosity. In this case, even a Galilean-invariant system with a single type of particle and a parabolic energy spectrum exhibits a temperature-dependent resistivity, as predicted by Gurzhi \cite{gurzhi}. However, the resistivity follows an inverse $T^{-2}$ dependence \cite{dejong, gusev1}, in contrast to the $T^{2}$ scaling observed in non-Galilean-invariant systems \cite{pal}. Negative magnetoresistance has been predicted in such narrow channels due to the sensitivity of viscosity to magnetic fields \cite{alekseev, scaffidi}. This effect has indeed been observed in GaAs two-dimensional electron systems, showing excellent agreement with theoretical predictions \cite{gusev1, raichev}. Furthermore, corrections to the Hall conductivity were also predicted and subsequently confirmed experimentally \cite{berdyugin, gusev2}.

Recent theoretical work has revealed a hydrodynamic origin for the giant magnetoresistance observed in compensated semimetals like charge-neutral graphene \cite{levchenko}. The theory shows that in the hydrodynamic regime, the Lorentz force distorts the electron fluid flow profile, generating a quadratic-in-field positive magnetoresistance at weak magnetic fields - providing the explanation for the giant magnetotransport effects near the CNP \cite{xin}.

A recent theoretical study of non-Galilean-invariant Fermi liquids without umklapp scattering \cite{maslov} has shown that electron-electron interactions do not generate magnetoresistance for isotropic spectra and suppress the usual $T^2$ scaling of magnetoconductivity for isotropic or convex 2D Fermi surfaces.

A system comprising two subbands with markedly distinct properties and carefully tuned parameters offers unique opportunities for studying how interparticle interactions affect magnetotransport. In our previous work \cite{gusev3}, we demonstrated that 6.3 nm HgTe quantum wells provide an interesting platform for examining interaction-dominated transport in 2D conductors. This hybrid band structure combines a single-valley Dirac cone near the zero-energy state with additional hole valleys having minima at finite wave vectors \cite{buttner, kozlov, gusev3, gusev4}. Crucially, when the chemical potential accesses these lateral heavy-hole bands in the valence band, Dirac holes experience strong scattering with heavy holes. These interaction processes break Galilean invariance, leading to a characteristic $T^{2}$ resistivity.

This work reports magnetotransport measurements, including magnetoresistance and the Hall effect, in gapless HgTe quantum wells with critical thicknesses of $d_c = 6.3-6.4$ nm (Fig.~\ref{fig1}(a)), emphasizing the regime of coexisting Dirac and parabolic holes. In this hybrid band system, the violation of Galilean invariance, coupled with inelastic hole-hole scattering, induces  corrections to both the classical Drude magnetoresistance and Hall resistivity. The corrections to magnetoresistivity exhibit a pronounced $T^2$ temperature dependence, in quantitative agreement with theoretical predictions for transport in massive-massless fermion systems.

\section{Electronic Structure of Gapless HgTe Quantum Wells}
\begin{figure*}
\includegraphics[width=16cm]{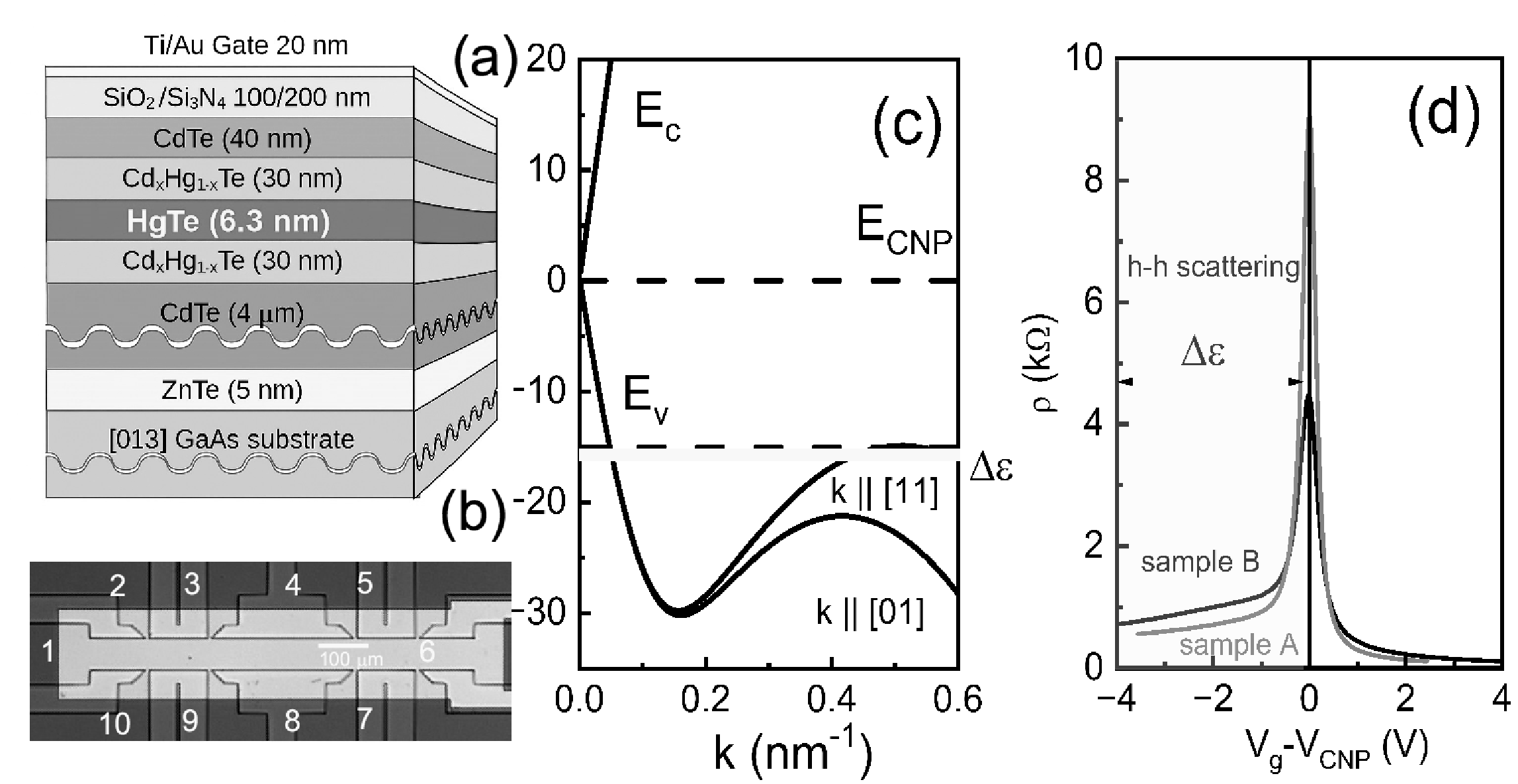}
\caption{ (a) Schematic of the transistor. (b) A top view of the sample. (c) Schematic representation of the energy spectrum of a 6.3-nm mercury telluride quantum well. (d) The resistivity of HgTe  quantum well as a function of the gate voltage for different samples.}
\label{fig1}
\end{figure*}

HgTe-based quantum wells have emerged as a versatile system for engineering two-dimensional electronic states with unique properties, including topological insulator phases and semimetallic behavior \cite{konig}. The electronic spectrum in these structures undergoes dramatic changes as a function of quantum well thickness, with the critical thickness range of $d_c=6.3-6.4$ nm producing a gapless spectrum characterized by a Dirac cone coexisting with parabolic bands \cite{gerchikov,kane,bernevig,bernevig2}.

The band structure of these quantum wells, shown in Fig.~\ref{fig1}(c), displays several distinctive features. Near the charge neutrality point, electrons and holes follow a linear dispersion relation $\varepsilon_e=\pm v|p|$ with a  high Fermi velocity $v=7\times 10^7 cm/s$. Below the Dirac point, the valence band forms a parabolic maximum with characteristic momentum  $p_0=m_{h} v_F=\sqrt{2m_{h}(\mu-\Delta)}$, where $m_h \approx0.15m_0$ represents the heavy hole effective mass, $\mu$ is the electrochemical potential, and $\Delta \approx 15$ meV denotes the indirect gap to the heavy hole band. For Dirac hole one may introduce the Dirac effective mass $m_d=\mu/v^2$.

Realistic samples exhibit inherent structural variations that significantly impact their electronic properties. Fluctuations in quantum well thickness ($\delta d \sim 0.1 nm$) during growth lead to spatial variations in both the band gap and charge neutrality point position. According to the topological network model described in Ref. \cite{gusev5}, these variations produce a density of states broadening on the order of 1-4 meV. Transport measurements reveal these effects through the asymmetric resistance peak at the charge neutrality point (Fig.~\ref{fig1}(d)), where the asymmetry stems from chemical potential pinning in the broadened heavy-hole band tails - a behavior markedly different from the symmetric resistance profile observed in graphene.

\begin{figure*}
\includegraphics[width=16cm]{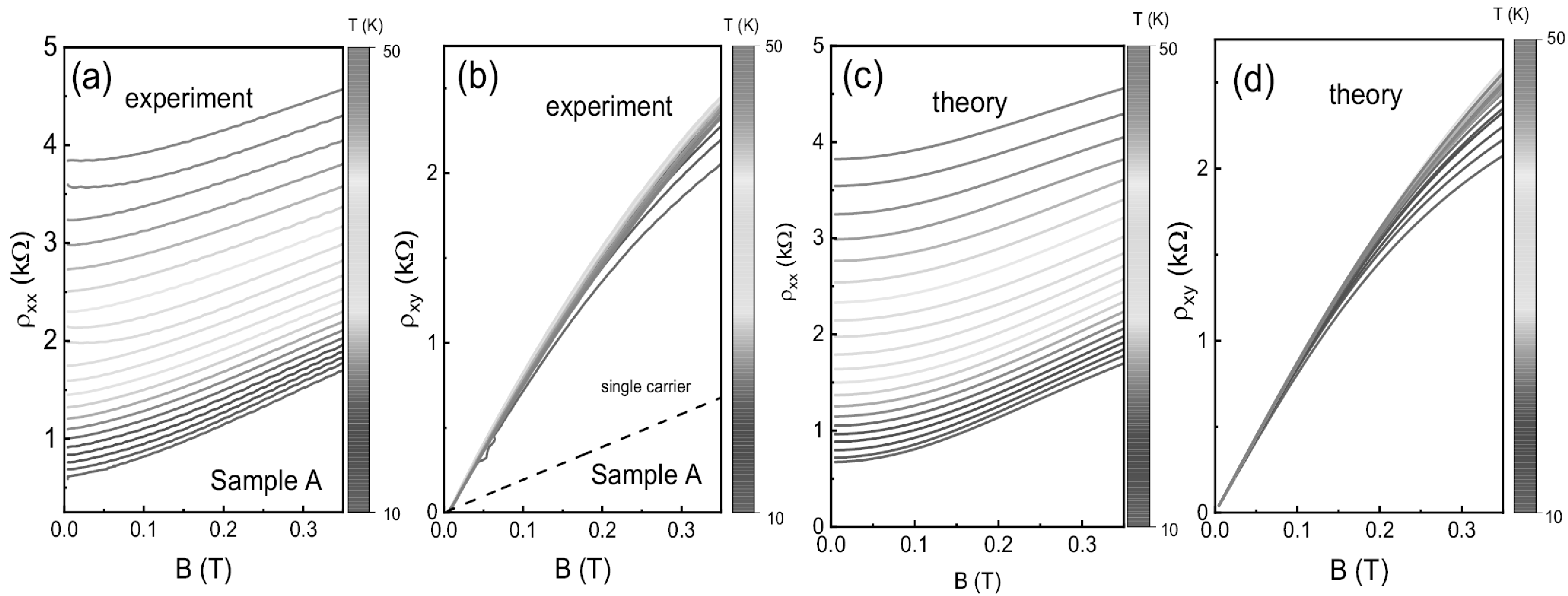}
\caption{ Gate voltage dependence of magnetoresistivity (a) and Hall effect (b) for different temperature  in the hole transport regime (sample A), $P_{total}=3.2\times 10^{11} cm^{-2}$ The gate voltage was varied in 2 K steps. The dashed line represents the Hall resistance calculated using the single-carrier model.
Panels (c) and (d) show the calculated B dependence of the magnetoresistivity and Hall effect at different temperature, respectively, based on Equations (4)–(10). The fitting parameters used in the calculations are presented in Fig. 3. }
\label{fig2}
\end{figure*}

\begin{figure}
\includegraphics[width=8cm]{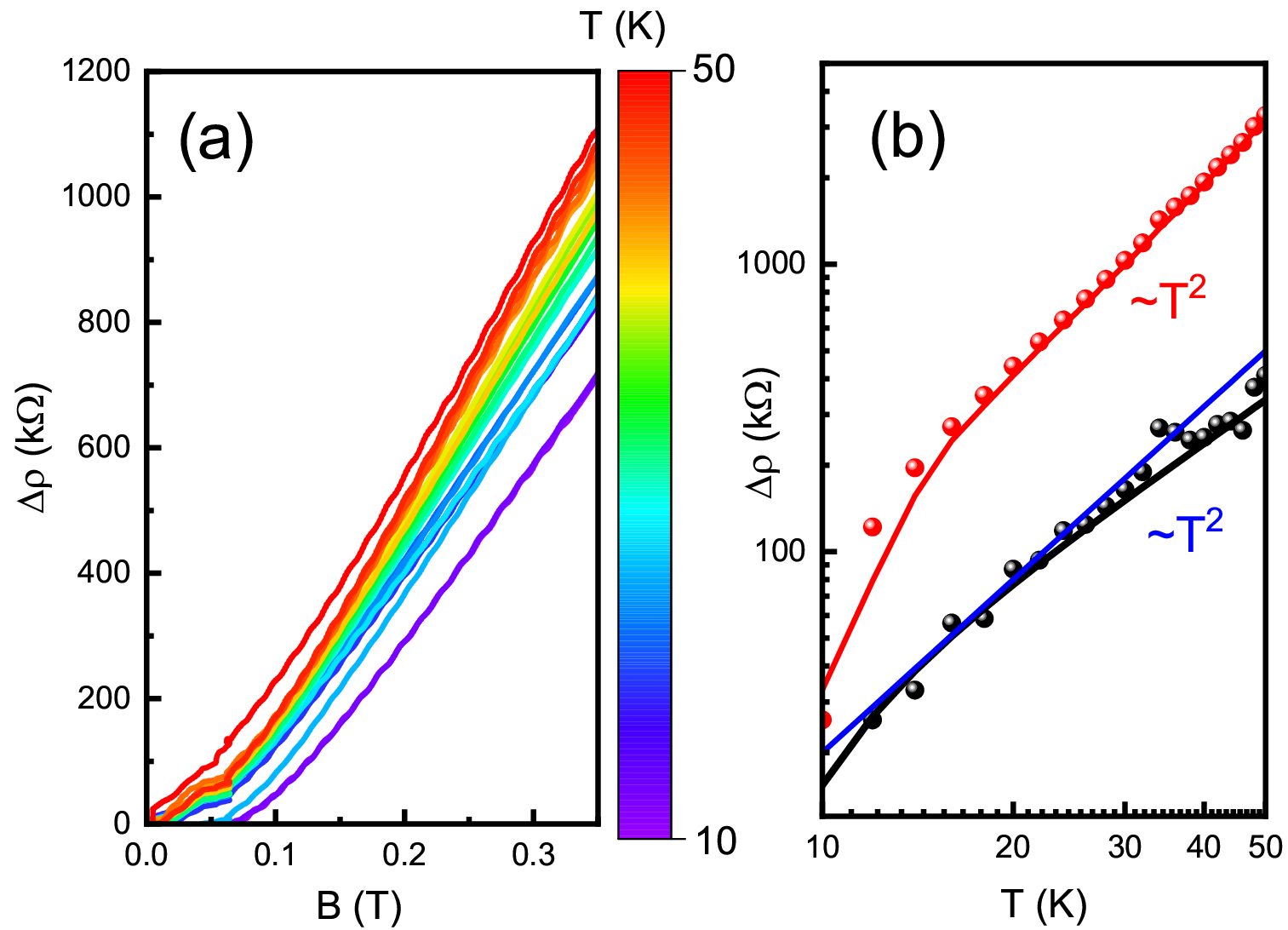}
\caption{ (a) The magnetoresistivity for different temperatures with substracted $\rho(B=0)$, measured at a fixed gate voltage
$V_{g} - V_{\text{CNP}} = 3.4\,\text{V}$ (sample A).
(b) Zero-field resistivity excess at $B = 0\,\text{T}$ (black circles), and magnetoresistivity excess defined as $\Delta\rho(10  K) - \Delta\rho(T)$ at $B = 0.35\,\text{T}$ (red circles), plotted as a function of temperature. Red and blue lines the calculated magnetoresistivity from Equation~(1) with parameters indicated in the text, blue line shows $T^2$ dependence.}
\label{fig3}
\end{figure}
\begin{figure*}
\includegraphics[width=16cm]{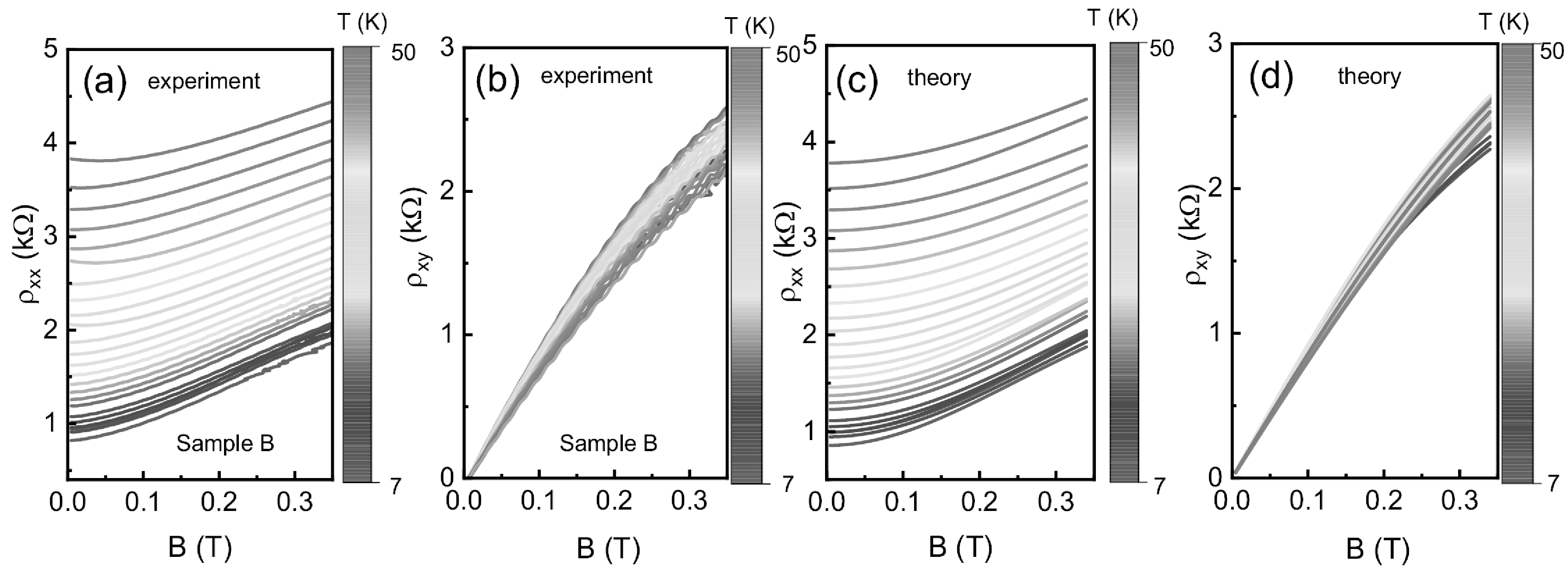}
\caption{ Gate voltage dependence of magnetoresistivity (a) and Hall effect (b) for different temperature  in the hole transport regime (sample B), $P_{total}=3.2\times 10^{11} cm^{-2}$ The gate voltage was varied in 2 K steps. The dashed line represents the Hall resistance calculated using the single-carrier model.
Panels (c) and (d) show the calculated B dependence of the magnetoresistivity and Hall effect at different temperature, respectively, based on Equations (4)–(10). The fitting parameters used in the calculations are presented in Fig. 5. }
\label{fig4}
\end{figure*}
\begin{figure}
\includegraphics[width=8cm]{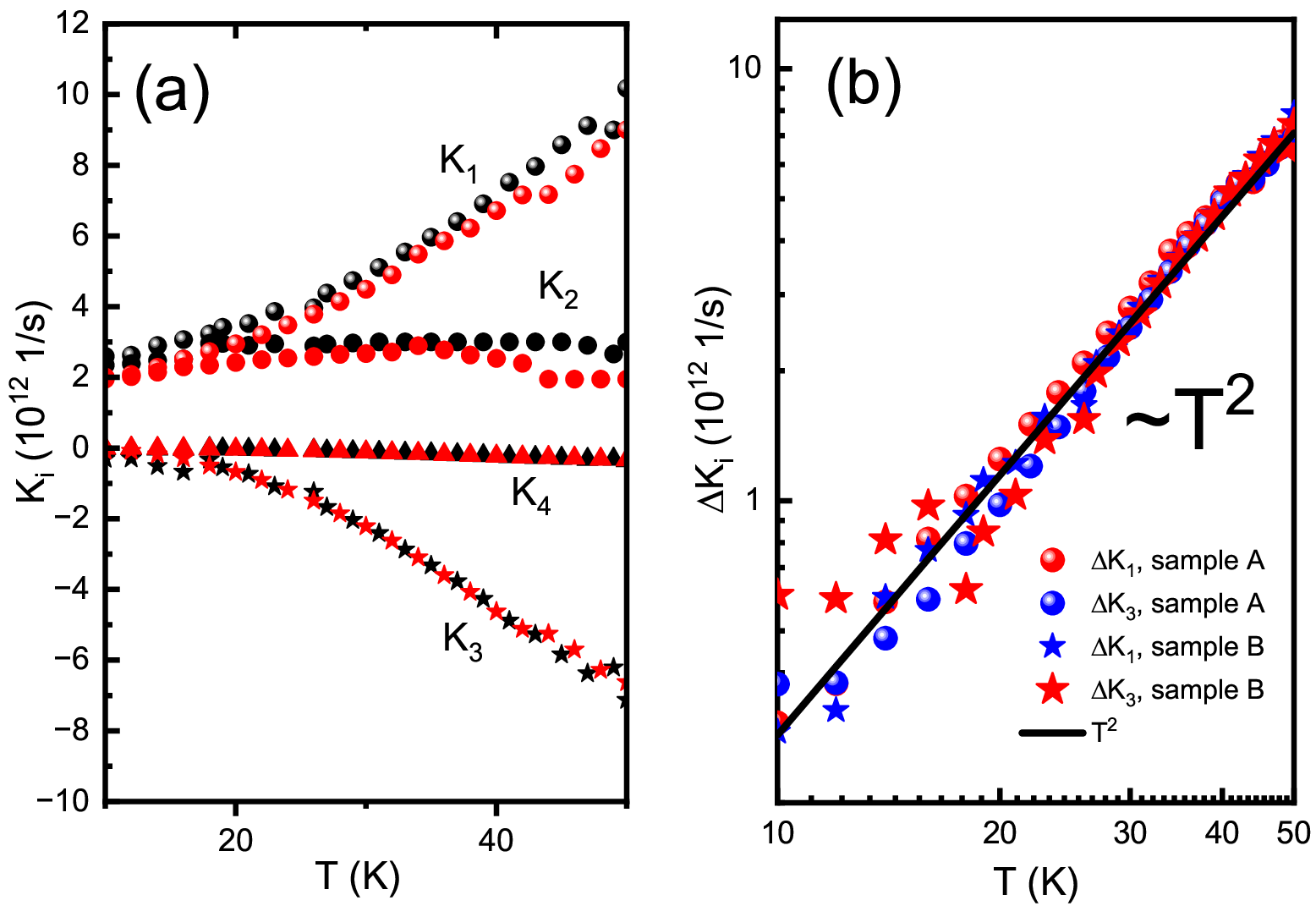}
\caption{ (a) The parameters $K_{i}$ of the scattering matrix ( eq.1) extracted from comparison with equations 1-10 for sample A (black marks) and sample  B(red marks).
(b) The parameter $\Delta K_{i}=K_{i}(T)-K_{i}(T=10K)$ of the scattering matrix ( eq.1) as a function of the temperature. Solid line is relaxation rate $\frac{1}{\tau_{dh}}\sim T^{2}$ calculated from equation~\ref{time}. }
\label{fig5}
\end{figure}
\section{Experimental Techniques and Device Characteristics}

The studied heterostructures consist of $ Cd_{0.65}Hg_{0.35}Te/HgTe$ quantum wells grown by molecular beam epitaxy on (013)-oriented GaAs substrates at temperatures between $160-200^{\circ}$ C \cite{mikhailov}. We investigated two representative devices (A and B) fabricated into Hall bar geometries with eight voltage probes distributed across three measurement segments of lengths 100, 250, and 100 ($\mu m$) and  width W=50 ($\mu m$) (Fig.~\ref{fig1}(b)) .

Device fabrication incorporated several key elements: a 200 nm $SiO_2$ gate dielectric layer, Ti/Au top gate electrodes, and In-based ohmic contacts formed through thermal annealing  (Fig.~\ref{fig1}(a)). The measurement approach employed low-frequency AC excitation (1-10 nA, 1-27 Hz) to minimize noise, with four-probe configurations used for both longitudinal $(R_{xx})$ and Hall $(R_{xy})$ measurements. The temperature-dependent transport properties were characterized from 4.2-50 K, with gate voltages limited to $\pm 8$ V by dielectric breakdown considerations.

The gate response showed a carrier density variation of $dn/dV_g=0.95\times10^{11} cm^{-2}V^{-1}$, enabling precise control of the Fermi level position relative to both the Dirac point and heavy-hole bands. Resistivity values were calculated using standard geometric corrections, with $\rho_{xx}
=(W/L)R_{xx}$ for the longitudinal component and $\rho_{xy}=R_{xy}$ for the Hall component. This comprehensive experimental approach provides simultaneous access to both Dirac and heavy-hole transport regimes while ensuring clean separation of longitudinal and Hall signals for detailed temperature-dependent analysis.
\section{Experimental results}
Fig.~\ref{fig1}(d) shows the gate voltage dependence of the resistance in a typical gapless HgTe quantum well device. As discussed in the previous section, our focus is on the gate voltage regime where hole–hole interactions dominate, highlighted by the yellow region in the figure. Although the gate voltage spans a broad range, from $-4 < V_g - V_{\mathrm{CNP}} < 0$, the corresponding change in chemical potential is confined to a narrow energy window, $-16.5\,\text{meV} < \Delta \epsilon < -15\,\text{meV}$, as shown by the yellow shading in Fig.~\ref{fig1}(c). This limited variation is due to the high density of heavy-hole states, which exceeds that of the Dirac holes by more than an order of magnitude.

As a result, the system enters a regime where degenerate Dirac holes coexist with non-degenerate heavy holes at elevated temperatures, satisfying the condition \( \mu - \Delta < kT < \mu \), known as the partially degenerate (PD) regime~\cite{gusev5}. In this regime, the resistivity exhibits a temperature dependence following a power law, \( \rho(T) \sim T^{\alpha} \) with \( \alpha \approx 3 \). This behavior is attributed to the scattering of non-degenerate heavy holes (obeying Boltzmann statistics) by degenerate Dirac holes. At lower temperatures, deviations from the \( T^3 \) dependence emerge, transitioning toward a \( T^2 \) dependence as the heavy holes become degenerate~\cite{gusev5}.

In Ref.~\cite{gusev5}, it was demonstrated that at higher temperatures, interaction-induced resistivity exceeds impurity-limited resistivity by a factor of 5–6. Based on this, it is reasonable to attribute the low-temperature magnetoresistance to the interplay between Dirac and heavy-hole subbands, governed by both interband and intraband scattering mechanisms.

Fig.~\ref{fig2}(a) displays the resistivity as a function of magnetic field at a fixed gate voltage
\( V_g - V_{\mathrm{CNP}} = 3.4\,\mathrm{V} \), corresponding to a total carrier density
\( P = 3.2 \times 10^{11}\,\mathrm{cm}^{-2} \), measured over a range of temperatures.
A pronounced positive magnetoresistance exceeding 100\% emerges on the hole side of the gate voltage sweep.
This magnetoresistance decreases slightly with increasing temperature, while the zero-field resistivity
exhibits a strong \(T^2\) dependence ( Fig.~\ref{fig3}(b), red cicles) consistent with the two-subband model and earlier reports~\cite{gusev3}.

Fig.~\ref{fig2}(b) presents the Hall resistance under the same experimental conditions.
The most remarkable observation is that the measured Hall coefficient,
\( R_{H} = \rho_{xy}/B \), is nearly an order of magnitude larger than the single-carrier
value \( 1/Pe \), where \( P \) is the total hole density. For reference, this expected value
is indicated by the dashed line in Fig.~\ref{fig2}(b). T
The strong deviation from single-carrier transport highlights the need for a two-subband model to capture the magnetotransport behavior~\cite{zaremba}.As shown in Fig.~\ref{fig2}(b), the temperature dependence of the Hall resistivity is nonmonotonic, in contrast to the magnetoresistivity. This distinction arises because the Hall resistivity is much more sensitive to the carrier density than to the scattering mechanisms. At finite temperature, the electron density is given by
\begin{equation}\label{density}
P_{d,e} = \int\limits_{-\infty}^{0} D_{\varepsilon}^{d,h}(\varepsilon)\,
\Bigl(1 - \bigl[1 + e^{(\varepsilon-\mu)/kT}\bigr]^{-1}\Bigr)\,d\varepsilon,
\end{equation}
where $D_{\varepsilon}^{d,h}$ denotes the density of states for Dirac and massive holes, respectively.  It is important to note that although the density in each subband can vary slightly with temperature at high $T$, the \textit{total} carrier density remains constrained by the gate voltage, such that $P_{d} + P_{h} = P \sim V_{g}$. Consequently, the chemical potential can be determined parametrically. In what follows, we compare our experimental results with theoretical predictions, taking into account the redistribution of carriers between the heavy and linear branches of the spectrum as $T$ increases. This redistribution may, however, lead to reduced agreement with theory, since additional parameters are introduced into the analysis.

To better visualize the temperature dependence, we first subtract the zero-field resistivity from each curve, and then compute the  magnetoresistivity at \( T = 10\,\mathrm{K} \) and that at other temperatures: $ \Delta\rho_{xx}(B)= \rho_{xx}(B)-\rho_{xx}(B=0)$. These results are plotted in Fig.~\ref{fig3}(a).  To further investigate the temperature dependence, we analyze the excess magnetoresistivity at a fixed magnetic field \( B = 0.35\,\mathrm{T} \) as a function of temperature: $\Delta \rho(T) = \Delta \rho(B = 0.35\,\mathrm{T}) - \Delta \rho(B = 0.35\,\mathrm{T}, T = 10\,\mathrm{K})$,
shown in Fig.~\ref{fig3}(b) (black circles). We assume here that the classical two-subband magnetoresistivity does not depend of the temperature, and all correction to this value are induced by interaction -induced corrections to the magnetoresistivity. For comparison, the zero-field resistivity excess is also plotted:
$\Delta \rho(B = 0) = \rho(T) - \rho(T = 10\,\mathrm{K})$,
depicted by black circles in the same figure. Both quantities, \( \Delta \rho(B = 0) \) and \( \Delta \rho(B = 0.35\,\mathrm{T}) \), exhibit a clear \( T^2 \) dependence, which is a hallmark of interaction-driven transport.
Fig.~\ref{fig4}(a,b) displays the resistivity and Hall effect as a function of magnetic field at a fixed gate voltage  \( V_g - V_{\mathrm{CNP}} = 3.4\,\mathrm{V} \), corresponding to a total carrier density  \( P = 3.2 \times 10^{11}\,\mathrm{cm}^{-2} \) for the sample B. The magnetoresistance and Hall effect exhibit similar behavior, confirming the robustness of the observed transport characteristics across different samples, despite  differences in absolute resistivity values, as also seen in Fig.~\ref{fig1}(d).

Therefore, the observed large positive magnetoresistance, strongly enhanced Hall response, and quadratic temperature dependence of the resistivity can be consistently explained by interaction-induced scattering processes between these subbands.

\section{Theory and Discussions}

A Boltzmann framework extended to multiple types of distinct charge carriers can be used to describe this behavior.  At low temperatures, the magnetoresistivity can be modeled using a two-subband theory that includes intersubband (intervalley) scattering, as developed by Zaremba \cite{zaremba}. In our hybrid band system, we analyze intervalley scattering, distinguishing "Dirac-heavy hole intervalley scattering" (between Dirac and heavy-hole carriers) from "heavy-hole intervalley scattering" (between heavy-hole valleys with non-zero k).  The scattering matrix K describes transitions between Dirac and heavy-hole states.
\begin{equation}
K_{ij} = \sum_{k \in \{d,hh\}} \Gamma_{ik}^0 \delta_{ij} - \Gamma_{ij}^1
\end{equation}

Scattering matrix elements from the matrix are:

\[
\mathbf{K} =
\begin{bmatrix}
K_1 & K_3 \\
K_4 & K_2
\end{bmatrix}
=
\begin{bmatrix}
\Gamma^{0}_{11} + \Gamma^{0}_{12} - \Gamma^{1}_{11} & -\Gamma^{1}_{12} \\[6pt]
-\Gamma^{1}_{21} & \Gamma^{0}_{21} + \Gamma^{0}_{22} - \Gamma^{1}_{22}
\end{bmatrix}
\]

where $\Gamma_{ij}^0$ is total scattering rate from band \(i\) to \(j\), such as $\Gamma_{ij}^1$ - momentum-weighted scattering rate (accounts for backscattering).

The eigenvalues  $\lambda_{1,2}$ of the scattering matrix \(\mathbf{K}\) are given by:
\begin{equation}
 \lambda_{1,2}= \frac{1}{2} \left( K_1 + K_2 \right)
\pm \frac{1}{2} \sqrt{\left( K_1 - K_2 \right)^2 + 4 K_3 K_4}
\end{equation}

In this framework, the magnetoresistivity exhibits a characteristic magnetic field dependence, with a parabolic behavior at low fields:
\begin{equation}
\rho_{xx}(B) = \frac{1}{\sigma_{0}}
\left(
  1 + \frac{\bar{\sigma}_{1}\bar{\sigma}_{2}(\gamma_{1}-\gamma_{2})^{2}}
  {\sigma_{0}^{2} + \left(\bar{\sigma}_{1}\gamma_{2} + \bar{\sigma}_{2}\gamma_{1}\right)^{2}}
\right),
\end{equation}

where total conductivity and partial conductivities can be written in the following form:
\[
\sigma_0 = \frac{e^2}{2 \pi}
\left( \frac{\alpha^2}{m_d \lambda_1} + \frac{\beta^2}{m_h \lambda_2} \right)
\equiv \bar{\sigma}_1 + \bar{\sigma}_2 .
\]
with $\gamma_i = \frac{\omega_i}{\lambda_i}$.
The corresponding cyclotron frequencies $\omega_i$ are $\omega_1 = eB/m_d$ for Dirac holes and $\omega_2 = eB/m_h$ for massive holes, reflecting their fundamentally different responses to magnetic fields. The effective densities $\alpha^2$ and $\beta^2$ are obtained from the following equations:
\begin{equation}
\alpha = \frac{k_h K_3 - k_d (\lambda_2 - K_1)}{\lambda_1 - \lambda_2},
\end{equation}

\begin{equation}
\beta = \frac{k_d (\lambda_1 - K_1) - k_h K_3}{\lambda_1 - \lambda_2}.
\end{equation}

where the Dirac and heavy-hole momenta are given by
\[
k_d = \sqrt{2\pi P_d}, \qquad k_h = \sqrt{\pi P_h},
\]
The magnetoresistivity $\rho_{xx}(B)$ can be expressed in the compact form:

\begin{equation}\label{magnetoresistance}
\rho_{xx}(B) = \frac{1}{\sigma_0} \left( 1 + \frac{a B^2}{1 + c B^2} \right),
\end{equation}

where the parameters $a$ and $c$ are defined as:

\begin{align}
a &= \frac{\bar{\sigma}_1 \bar{\sigma}_2 (\mu_1 - \mu_2)^2}{\sigma_0^2}, \\
c &= \frac{(\bar{\sigma}_1 \mu_2 + \bar{\sigma}_2 \mu_1)^2}{\sigma_0^2},
\end{align}

with the auxiliary quantities: $\mu_1 = \frac{e}{m_d \lambda_1}$, $\mu_2 = \frac{e}{m_h \lambda_2}$.

 The Hall component of the resistivity exhibits a linear magnetic field dependence at low fields:
\begin{equation}
\rho_{yx} = -\frac{\gamma_{1}\bar{\sigma}_{1} + \gamma_{2}\bar{\sigma}_{2} + \gamma_{1}\gamma_{2}(\gamma_{1}\bar{\sigma}_{2} + \gamma_{2}\bar{\sigma}_{1})}{(\bar{\sigma}_{1} + \bar{\sigma}_{2})^{2} + (\gamma_{1}\bar{\sigma}_{2} + \gamma_{2}\bar{\sigma}_{1})^{2}}
\end{equation}

The scattering process from Dirac to heavy-hole valleys ($d \rightarrow hh$) is primarily governed by short-range defects (such as vacancies or impurities), since it requires a large momentum transfer.

In contrast, the reverse process, corresponding to heavy holes scattering into Dirac states ($hh \rightarrow d$), is significantly less probable. This suppression arises from phase-space restrictions, namely the limited availability of final Dirac states at low energies.

The intervalley scattering rate from Dirac to heavy holes, $\Gamma_{1,2}=\Gamma_{d,hh}$, can be approximated as:
\begin{equation}
\Gamma_{d,hh} \propto
\frac{2\pi}{\hslash} n_{\text{imp}} |V_0|^2
\left( \frac{m_{h}}{\pi \hslash^2} \right)
\end{equation}
where $n_{\text{imp}}$ is the impurity concentration, $V_0$ is the scattering potential strength and $m_{h}$ is the heavy-hole effective mass.

For the reverse channel, the scattering rate is instead proportional to the energy-dependent density of states (DOS) of Dirac holes:
\begin{equation}
\Gamma_{2,1}=\Gamma_{hh,d}(\epsilon) \propto N_d(\epsilon) = \frac{\epsilon}{\pi \hslash^2 v^2}.
\end{equation}
This results in a pronounced asymmetry: the $d \rightarrow hh$ scattering rate increases with the Dirac-hole Fermi energy $\epsilon$, reflecting the larger number of available initial states. In contrast, the $hh \rightarrow d$ channel remains suppressed at low $\epsilon$ due to the limited availability of final Dirac states.

This asymmetry allows us to introduce an additional constraint. Specifically, we impose strong backscattering in the intervalley parameters, assuming
\begin{equation}
\frac{K_{3}}{K_{4}} = \frac{\Gamma_{hh,d}}{\Gamma_{d,hh}} = \frac{\epsilon}{m_{h} v^2} \approx 0.04.
\end{equation}

Next, we fit Equations (2) and (9) to describe the zero-field resistivity, magnetoresistivity, and Hall effect. The fitting procedure involves three parameters: $K_1$, $K_2$, and $K_3$. The hole densities are calculated from the total carrier density, which is determined based on the gate voltage dependence and the density of states in the valence band, as described in Ref.~\cite{gusev3}. The results of the fitting are presented in Fig.~\ref{fig2}(c,d) for the sample A and in Fig.~\ref{fig4}(c,d) for the sample B. We find excellent agreement with the theoretical model of Ref.~\cite{zaremba}, as modified here for hybrid band systems, across a broad range of magnetic fields and temperatures. By comparing both samples with the theoretical model, we also confirm that the observed behavior is reproducible across different devices.  Strikingly, we are able to simultaneously describe several key observations, including a tenfold enhancement of the Hall resistance, a large positive magnetoresistance, and a $T^2$ dependence of the zero-field resistivity. In contrast, comparison with a simplified two-subband model \cite{ziman}, which does not account for intravalley scattering, fails to reproduce our results for any choice of adjustable parameters.

The variation of the parameters $K_1$, $K_2$, $K_3$, and $K_4$ with temperature is presented in Fig.~\ref{fig5}(a). Among them, $K_1$ and $K_3$ exhibit a noticeable temperature dependence, whereas $K_2$ and $K_4$ remain nearly independent of temperature. To analyze the functional dependence of the parameters, it is useful to subtract the temperature-independent contributions from $K_1$ and $K_3$, and evaluate the excess relaxation rates defined as $\Delta K_i = K_i(T) - K_i(T = 10,\text{K})$. Fig.~\ref{fig5}(b) shows the excess values $\Delta K_i$ as a function of temperature for two devices. Remarkably, all dependencies converge to a unique $T^2$ behavior, indicating a universal mechanism governing the scattering-rate relaxations.  To account for the temperature dependence of the parameters, and consequently the magnetoresistivity and Hall effect at elevated temperatures, we further refine the previously discussed model by incorporating inelastic scattering processes between Dirac fermions and heavy holes. Specifically, these inelastic processes introduce additional terms to the scattering matrix $\mathbf{K}$ that capture the enhanced intervalley scattering rates at higher temperatures. The modified scattering matrix is given by
\begin{equation}
K = \begin{bmatrix}
K_1^*  &K_3^*  \\
K_4^* & K_2^*
\end{bmatrix}=\begin{bmatrix}
K_1 + \frac{1}{\tau_{dh}} &K_3 -\frac{1}{\tau_{dh}} \\
K_4-\frac{1}{\tau_{hd}} & K_2 + \frac{1}{\tau_{hd}}
\end{bmatrix}
\end{equation}
where $1/\tau_{dh}$ represents the inelastic scattering rate from Dirac holes to heavy holes, and $1/\tau_{hd}$ denotes the reverse process from heavy holes to Dirac holes. These rates are expected to exhibit a quadratic temperature dependence, such as $1/\tau_{dh} \propto T^2$, consistent with the observed $\Delta K_i \propto T^2$ behavior in Fig.~\ref{fig5}(b), arising from mechanisms of the interparticle scattering in the hybrid band system. This modification preserves the overall structure of the two-subband theory while effectively increasing the diagonal elements and reducing the off-diagonal coupling for the Dirac-heavy hole transitions, leading to a suppression of coherent backscattering and a corresponding evolution in the magnetotransport properties as temperature rises. With this refinement, the eigenvalues and effective densities can be recalculated, yielding improved fits to the experimental data across the full temperature range.
The scattering times are described by \cite{gusev3, levin}:
\begin{gather}\label{time}
\frac{1}{\tau_{dh}}=A\frac{m_h (kT)^2|U_0|^2}{3\pi\hslash^5vv_F}\sim T^{2},\\\nonumber
\frac{1}{\tau_{hd}}=\frac{(\mu/v^2) (kT)^2|U_0|^2}{3\pi \hslash^5vv_F}\frac{k_0^2}{p_0^2}\sim T^{2}.
\end{gather}
 Here, $k_0$ is defined by $v k_0 = \mu$, and $p_0 = m_h v_F = \sqrt{2 m_h (\mu - \Delta)}$ for degenerate Dirac and heavy-hole gases. The coefficient $A$ is a numerical factor that depends on the details of the Coulomb interactions.

The contact interaction parameter $U_0$ can be estimated as over-screened Coulomb interaction, $U_0=2\pi e^2/\varepsilon q_s$, where $q_s$ is a screening wave vector.
It is worth noting that this prescription applies to strongly degenerate Fermi liquids. Assuming that the main contribution to the screening is determined by the heavy holes due to its high density of states, the screening wave vector can be written as $q_s=m_he^2/(\epsilon\hslash^2)$, where $\epsilon=10$ is the dielectric constant of the material.

The equation ~\eqref{time} considering the constraint that interactions between holes conserve the over-all momentum density $m_dP_d\tau_{dh}=m_hP_h\tau_{hd}$. It follows from this equation that $\tau_{dh} \gg \tau_{hd}$, given that other conditions are also satisfied, namely $P_h > P_d$ and $m_h \gg m_d$.  This constraint explains our observations and the temperature dependencies shown in Fig.~\ref{fig5}(a,b). According to the matrix in Eq.~13, the excess relaxation rate $\Delta K_1 = 1/\tau_{dh} \sim T^2$, whereas $\Delta K_2\sim \Delta K_4 \sim 1/\tau_{hd} \ll K_1$ are very small and nearly temperature-independent, since they are negligible compared to the other parameters $K_i$.

In Fig.~\ref{fig5}, we compare the calculated results with the experimental $1/\tau_{dh}$ dependencies using Eq.~\eqref{time}, and plot both the experimentally measured and theoretically predicted excess relaxation rates $\Delta K_i$ for two samples. To compare with theory, we performed a fitting analysis of the temperature-dependent data, as shown in Fig.~\ref{fig5}, using a single adjustable parameter that accounts for the strength of the interaction between Dirac fermions and heavy holes $A=2.8$. One would expect, however, that $A\approx 1$.
First, for a quantum well of finite width, the Coulomb matrix element acquires a form factor $F(q)$, and screening becomes $q$-dependent. Accordingly, the effective interaction is more accurately written as
\[
U(q,\omega)\sim \frac{2\pi e^2}{q}\,\frac{|F(q)|^2}{\epsilon(q,\omega)} ,
\]
rather than a constant $U_0$. This is a standard result for 2D carriers in semiconductor quantum wells (see Ref.~\cite{ando}). In our experiment, the combined effect of the form factor and the $q$-dependent polarizability can renormalize the effective interaction by a factor of order unity (typically $\sim 2$--3), consistent with the fitted value $A\simeq 2.8$.
Second, the relevant processes occur at a finite transferred frequency $\omega\sim T/\hslash$, where screening is generally weaker than in the static limit, which increases $|U|^2$ and thus enhances the inelastic rates. Importantly, for Hg(Cd)Te/HgTe quantum wells, the static and dynamical polarizability within the random phase approximation (RPA) has been analyzed in the literature, e.g., in Ref.~\cite{juergens}, and related dynamical-screening calculations for HgTe QWs can be found in Ref.~\cite{melezhik}. These works show that $\epsilon(q,\omega)$ in HgTe QWs can deviate substantially from the simplest Thomas--Fermi ``contact'' approximation, especially when band non-parabolicity/Dirac-like features and finite $(q,\omega)$ are relevant. A full quantitative evaluation of $A$ would require calculating the mixed Dirac--heavy-hole scattering rate using $U(q,\omega)$ together with the HgTe QW polarization function and finite-thickness form factors, which is beyond the scope of the present transport-focused paper. Instead, following the general theory of relaxation in 2D Fermi systems with screened Coulomb interactions (see Ref.~\cite{dmitriev}), where the electron--electron scattering rate is highly sensitive to the interaction strength and screening details, we absorb these corrections into a single renormalization factor $A$.

The scattering rates $1/\tau_{dh}$ predominantly determine the behavior of $\Delta K_i$ at high temperatures.

It can be readily shown that, in the high magnetic field limit, the saturation of the magnetoresistance is given by
\begin{equation}
\rho_{xx}(B) \to \frac{1}{\sigma_0} \left(1 + \frac{a}{c}\right).
\end{equation}

Using the conditions $\frac{1}{\tau_{dh}} = \alpha T^2 \gg \frac{1}{\tau_{hd}} \gg K_1, K_3 \gg K_4, \quad K_1 \sim K_2, \quad m_d \ll m_h, \quad P_h \gg P_d \ (\text{implying } k_h \gg k_d)$, we can simplify the coefficients as $a \approx \frac{\bar{\sigma}_1 \mu_2^2}{\bar{\sigma}_2}, \quad c \approx 4 \mu_1^2$, which leads to
\begin{equation}
\frac{a}{c} \approx \frac{\bar{\sigma}_1 \mu_2^2 / \bar{\sigma}_2}{4 \mu_1^2} = \frac{1}{4} \left( \frac{m_d}{m_h} \right) \left( \frac{\lambda_1}{K_2} \right).
\end{equation}

Since $\lambda_1 \approx \alpha T^2$, we find $\frac{a}{c} \propto T^2$. In the low-temperature limit, where inelastic contributions vanish  and the base scattering parameters dominate, $a/c$ becomes approximately constant and temperature-independent. Therefore, the difference in the saturation values between the low-$T$ (constant) and high-$T$ ($\propto T^2$) limits scales as $T^2$, as illustrated in Fig.~\ref{fig3}(b). We also directly calculated this difference $\Delta \rho(T) = \Delta \rho(B = 0.35\,\mathrm{T}) - \Delta \rho(B = 0.35\,\mathrm{T}, T = 10\,\mathrm{K})$ using Eqs.~1--13. The results, shown in Fig.~\ref{fig3}(b), demonstrate good agreement with the experimental data.

Note that Eq.~(7) yields $\rho_{xx}\propto B^{2}$ at low fields and saturation at high fields (Eq.~(16)). In our field range, the data correspond to the crossover regime $cB^{2}\sim 1$, where the theoretical dependence can appear nearly linear over a limited interval of $B$, although the asymptotic high-field behavior is saturating.

Therefore our modified model, incorporating inelastic hole-hole scattering between Dirac and heavy-hole carriers, successfully describes the full spectrum of transport behavior at elevated temperatures, encompassing the three fundamental measurable quantities: zero-field resistivity, magnetoresistivity, and the Hall effect. By accounting for the temperature-dependent enhancements in the scattering matrix, we achieve excellent agreement with experimental data across a wide range of conditions and devices. These results underscore the critical role of hole-hole interactions in governing magnetotransport in hybrid band systems, revealing that both the excess relaxation rates $\Delta K_i$ and the associated scattering processes exhibit a pronounced $T^2$ dependence, thereby highlighting a universal mechanism driven by thermally activated interparticle scattering.

\section{Conclusion}
In the interaction-dominant regime, our study of the 6.3 nm gapless HgTe quantum well (QW) reveals that this two-dimensional hybrid band system, characterized by coexisting linear (Dirac-like) and parabolic hole energy bands, displays temperature dependent Hall resistance and substantial positive magnetoresistance. These phenomena are predominantly governed by the interplay of high-mobility Dirac holes and inelastic hole-hole scattering processes, which become increasingly significant at elevated temperatures. Employing a refined classical two-subband model that integrates intervalley scattering with temperature-dependent inelastic contributions—manifesting as quadratic \(T^2\) dependencies in scattering rates—we achieve precise descriptions of the magnetoresistivity and Hall effect over extensive ranges of magnetic fields, carrier densities, and temperatures.

This investigation into the HgTe QW provides a solid foundation for comprehending mixed carrier magnetotransport in hybrid band systems under interaction-dominant conditions, where the enhanced two-subband model serves as an effective instrument for elucidating intricate dynamics driven by interparticle interactions. Applying this extended framework to other hybrid band systems, including three-dimensional topological insulators and Weyl semimetals, holds promise for elucidating the roles of linear and parabolic dispersions, intersubband scattering, topological influences, and thermally activated interactions in shaping transport behaviors across varied platforms. Through adaptations of the multi-subband model to accommodate system-specific intricacies, such research may unveil innovative transport effects, deepen insights into hybrid band systems, and facilitate the development of next-generation quantum technologies.

\section{Data Availability}
The datasets generated and  analysed during the current study are available in Zenodo repository. https://10.5281/zenodo.17279292  \cite{zenodo}

\section*{Acknowledgements}
The authors thank V.~M.~Kovalev for valuable discussions and correspondence.

\section*{Funding}
This work was supported by the São Paulo Research Foundation (FAPESP) Grants No.~2019/16736-2 and No.~2021/12470-8,
and by the National Council for Scientific and Technological Development (CNPq).
The growth of HgTe quantum wells and preliminary transport measurements were supported by the Russian Science Foundation (Grant No.~23-72-30003).

\end{document}